\documentstyle[prb,epsfig,aps]{revtex}
\begin{document}
\draft
\title{From bi--layer to tri--layer Fe nanoislands on Cu$_{3}$Au(001).}

\author{A. Verdini, 
L. Floreano,\footnote{Corresponding Author:
Luca Floreano,
Surface Division,
Laboratorio TASC-INFM,
Basovizza, SS14 Km 163.5
I-34012 Trieste, Italy.
Fax: +39-040-226767;
E-mail: floreano@sci.area.trieste.it} 
F. Bruno,
D. Cvetko,$^{a}$ 
A. Morgante,$^{b}$ 
}
\address{
Laboratorio TASC dell'Istituto Nazionale per la Fisica della
Materia, S.S.14 Km.163.5, Basovizza, I-34012 Trieste, Italy\\
$^{a}$ also at: J. Stefan Institute,
University of Ljubljana, Slovenia, and Sincrotrone Trieste,
Italy.\\
$^{b}$ also at: Dipartimento di Fisica dell'Universit\`a di Trieste,
Italy.
}
\author{
F. Bisio, 
S. Terreni
and M. Canepa}

\address{
INFM and Dipartimento di Fisica dell'Universit\`a di Genova,
Italy.
}

\narrowtext
\onecolumn

\maketitle

\begin{abstract}

Islands of 1-2~nm lateral size and double layer height 
are formed when 1 monolayer (ML) of Fe is deposited on 
Cu$_{3}$Au(001) at low temperature.
We used the  PhotoElectron Diffraction (PED) technique to investigate 
the atomic structure and chemical composition of these 
nanoislands just after the deposition at 140~K and after annealing at 400~K. 
We show that only bi--layer islands are formed at low temperature, 
without any surface segregation. After annealing, the Fe atoms are 
re-aggregated to form mainly tri-layer islands. Surface 
segregation is shown to be inhibited also after the annealing process.
The implications for the film 
magnetic properties and the growth model are discussed.

\end{abstract}
\vspace{1cm}


\twocolumn
\narrowtext
%\onecolumn

The properties of nanometers size clusters of magnetic materials are 
subject of extended studies for their relevance in the miniaturization 
of memory storage devices. Both the reduced dimensionality and system size 
are responsible for magnetic behavior different from that observed in 
bulk materials. 
Self assembly on suitably chosen substrates 
is a well exploited root to control the structure and morphology, 
hence magnetization, of metal films. 
Much attention has been devoted to 
ultrathin Fe films, due to the possibility of stabilizing Fe in 
magnetic phases different from the bcc bulk ones, i.e. either 
superferromagnetic or antiferromagnetic, or even 
nonmagnetic.\cite{moroni,moruzzi}
In particular, the Cu$_{3}$Au(001) surface has been recently singled 
out as a good template to grow high spin Fe 
phases,\cite{rochow,lin,feldmann} 
due to the close matching between the 
Cu$_{3}$Au lattice constant (3.75~\AA) and the equilibrium lattice 
constant for fcc ferromagnetic Fe (3.65~\AA).

In fact, fcc Fe films on Cu$_{3}$Au(001) are obtained with an 
increasing degree of 
tetragonal distortion as the thickness is increased, until a bcc--like
phase is finally recovered.\cite{terreni,luches,lin2,schirmer}
Previous research mostly dealt with the spin reorientation transition 
of the surface magnetization from out-of-plane to in-plane and its 
possible correlation with the structural transition from fcc to 
bcc.\cite{lin,feldmann}
Here we will focus on the initial stages of growth, where Fe is 
expected to be pseudomorphic to the substrate, thus yielding the largest 
volume of its lattice cell, hence the highest magnetic 
moment.\cite{moroni}
While growth proceeds almost layer by layer at room temperature 
(RT),\cite{lin2} with a small amount of Au 
segregation,\cite{luches,schirmer}
the formation of uniform size (on the nanometer scale length) 
flat bi--layer islands
was recently observed by He atom 
scattering (HAS) after deposition of 1~ML of Fe at 140~K.\cite{canepa}
Bi--layer nanoislands of this kind were also reported for a few 
metal on metal systems in 
the first monolayer range.\cite{chambliss,giergiel,figuera} 
Several mechanisms have been suggested to drive the 
bi--layer growth such as magnetostriction effects,\cite{giergiel} 
combination of strain and electron confinement,\cite{canepa} and
a combination of strain and exchange processes.\cite{gomez}
A better knowledge of the inner structure and chemical composition of 
these nanoislands would be helpful to understand the driving force 
for the bi--layer growth. 
Up to now these systems have been investigated by techniques such as 
HAS and scanning tunneling microscopy (STM) which only probe the outer 
surface valence charge density. 

In the present work, we have exploited the experimental techniques 
available at the ALOISA beamline\cite{floreano} of the Elettra Synchrotron 
(Trieste, Italy) for a structural and chemical characterization of the islands 
formed after deposition of 1~ML of Fe at 140~K, and subsequent 
annealing to 400~K. 
Surface X-ray diffraction (XRD), X-ray specular reflectivity (XRR) and PED were
previously used for a complete characterization of the Fe/Cu$_{3}$Au(001) 
system in a higher thickness range (3-36~\AA).\cite{terreni}
The same sample of Refs.~\cite{terreni,luches,canepa} has been used, 
further details about sample 
preparation are given elsewhere.\cite{terreni,mannori} 
After calibration, XRR allowed us to monitor the deposition of a 
single Fe ML with a high level of precision and reproducibility. 
PED patterns have been taken to determine the structure of Fe islands 
and to check the occurence of Au segregation at the Fe island surface.  
We measured the photoelectrons of the Fe 2p$_{3/2}$ core--level 
at a Kinetic Energy of 231~eV. 
The PED polar scans have been measured by collecting the photoemission 
signal as a function of the emission angle $\theta$,  by rotating the electron 
analyzer in the scattering plane. The grazing angle 
was kept fixed at 4.5$^{\circ}$, with the polarisation in transverse magnetic 
condition and the surface oriented with the $\langle 100 \rangle$ 
direction in the scattering plane.

\begin{figure}[tbp]
\includegraphics[width=.45\textwidth]{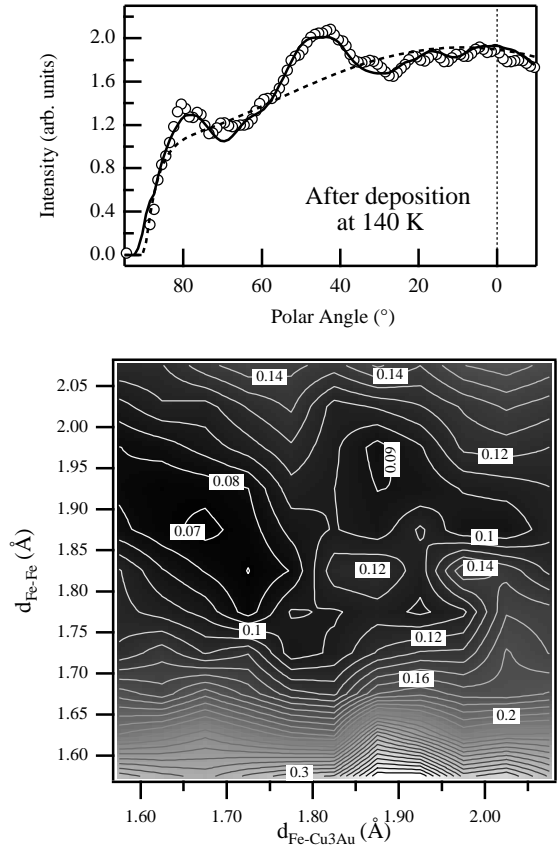}
\caption{Upper panel: PED polar scan taken along 
the $\langle 100 \rangle$ symmetry 
direction after the deposition at 140 K (open circles). 
The best--fit model (full line) of the bi--layer is reported, together with 
the best ISO background (dashed line). The surface normal is 
indicated by the vertical dotted line at 0$^{\circ}$.
Lower panel: R-factor analysis of the PED simulation for different 
values of the vertical spacing between the two Fe layers $d_{Fe-Fe}$ 
and the Fe-substrate layers  $d_{Fe-Cu3Au}$.}
\label{fig1}
\end{figure}

The photoelectron angular modulation is given by the superposition of 
two contributes: i) the diffractive atomic--position--dependent anisotropy term 
$\chi$--function, which has been determined by multiple scattering 
calculations (MSCD code\cite{MSCD} with hemispherical 
clusters of radius~=~9.5~\AA\ 
and about 170 atoms, Multiple--Scattering--order~=~6, Rehr--Albers--Order~=~2, 
Debye Temperature~=~220~K and inner potential V$_{0}$~=~10~eV); 
ii) the isotropic background $ISO (\theta)$
arising from the emission matrix element, sample attenuation, 
surface roughness, sample illumination and analyzer response, which have been 
described by an analytical expression.\cite{terreni}   Both 
contributes have been compared to simulations by means of a 
reliability R-factor analysis.

\begin{figure}[tbp]
\includegraphics[width=.45\textwidth]{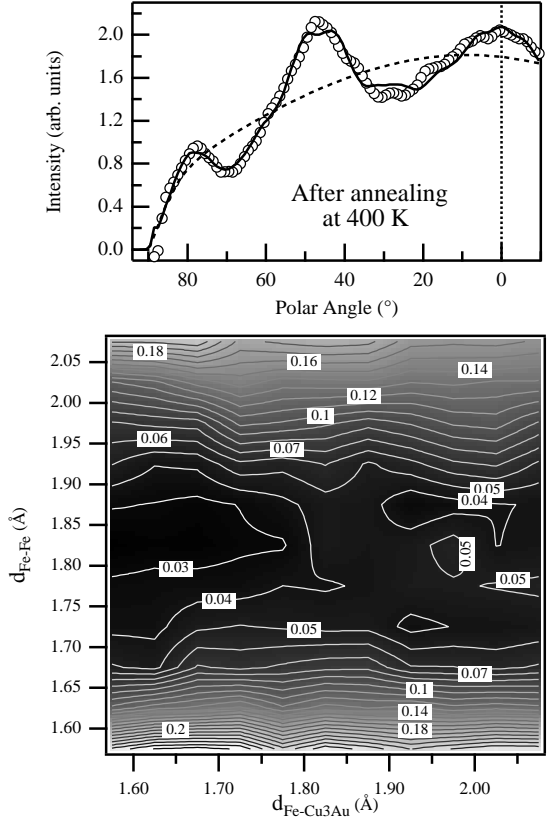}
\caption{Upper panel: PED polar scan taken along 
the $\langle 100 \rangle$ symmetry 
direction after annealing at 400 K and cooling to 140~K (open circles). 
The best--fit model (full line) of the tri--layer is reported, 
together with the best ISO background (dashed line).
The surface normal is 
indicated by the vertical dotted line at 0$^{\circ}$.
Lower panel: R-factor analysis of the PED simulation for different 
values of the vertical spacing between the Fe layers $d_{Fe-Fe}$ 
and the Fe-substrate layers  $d_{Fe-Cu3Au}$. Here the Fe interlayer 
spacing has been kept equal for both the first to second and the second to 
third Fe layer spacing. }
\label{fig2}
\end{figure}

A qualitative inspection of the polar scan taken after deposition at 
140~K indicates the formation of a
double layer fcc--like structure (see Fig.~\ref{fig1}, upper panel). 
In fact, the anisotropy $\chi$-function along the  $\langle 100 \rangle$ 
direction is characterized by an 
intense peak at  about 
45$^{\circ}$, which is the fingerprint of an fcc structure. 
At the same time, no diffraction 
features are observed along the surface normal; this
can be ascribed to the fact that the third layer of the fcc structure 
is not formed yet. 
 The polar scan, taken after annealing 
the film at 400~K for 5 minutes and cooling down to 140~K, is reported in 
the upper panel of Fig.~\ref{fig2}. 
The experimental data display an overall increased anisotropy ; 
in particular the increase of the anisotropy at 
the surface normal and at 45$^{\circ}$ is a clear hint of the formation of the third 
layer in the fcc structure. 

The best--fits, shown 
in Figs.~\ref{fig1} and \ref{fig2}, have been obtained by keeping the lateral lattice 
parameter fixed to the substrate value of 2.65~\AA~ 
(i.e. the half-diagonal of the Cu$_{3}$Au(100) lattice constant), 
according to the XRD indication from 
ref.~\cite{terreni}. The separation between the Fe--Fe and Fe--substrate 
layers ($d_{Fe-Fe}$ and $d_{Fe-Cu3Au}$, respectively) have been used as fitting 
parameters. 
In the case of bi--layer islands, we found a certain degree of 
correlation between $d_{Fe-Fe}$ and $d_{Fe-Cu3Au}$, thus yielding a 
good estimate of the total island height $d_{Fe-Fe} + d_{Fe-Cu3Au} 
\sim$3.55~\AA, but a rather large uncertainty in the individual 
parameters $d_{Fe-Fe}$=1.87$\pm$0.1~\AA~ and 
$d_{Fe-Cu3Au}$=1.67$\pm$0.1~\AA. In the case of tri--layer islands, 
the spacing between Fe layers was found to be 
$d_{Fe-Fe}$=1.83$\pm$0.05~\AA~ for both the topmost and inner Fe 
layers (i.e. no surface relaxation was detected).
The latter value corresponds to the vertical lattice 
spacing of the fcc substrate, in full agreement with the value 
previously obtained for an Fe film of 3~\AA~ nominal thickness (about 2~ML). 
\cite{terreni} The Fe--substrate separation is not equally well 
determined, even if it points to a contracted value 
$d_{Fe-Cu3Au}$=1.65$\pm$0.15~\AA~ for the tri--layer islands too. 

\begin{figure}[tbp]
\begin{center} 
\includegraphics[width=.45\textwidth]{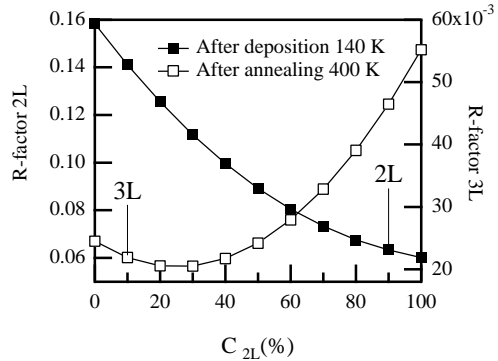}
\end{center}
\caption{R-factor analysis for evaluating the concentration of bi- and 
tri--layer Fe islands. The comparison is made with a 
linear combination of double and triple layer Fe 
models for the low temperature deposition 
(filled squares) and after annealing at 400~K (open squares). 
The concentration $C_{2L}$ is the weight in \% of the 
double layer model.}
\label{fig3}
\end{figure}

Once obtained the film structure,
the height distribution of the islands for each film has been 
determined by fitting the polar scans to a linear combination of the 
$\chi$-functions calculated for the double and triple layer films:

\begin{equation}
\chi=C\cdot \chi_{2ML}+(1-C)\cdot \chi_{3ML},
\label{eq2}
\end{equation}

where $C=C_{2L}$ plays the role of "concentration" of the double layer film.  
From the R-factor analysis in Fig. \ref{fig3}, we can say that 
no tri--layer islands are formed after the 
deposition at LT (as expected by the inspection of the polar scan in 
the upper panel of Fig.~\ref{fig1}). 
The comparison to the experimental data taken after the annealing process 
(open squares) presents a minimum of the R--factor at about $C_{2L}$ = 25\%. 
Therefore, after the annealing process there is a majority amount ($\sim$75\%) of 
tri-layer islands. The results for both as--deposited and annealed 
films are in full agreement with previous HAS analysis.\cite{canepa}
The eventual presence of single layer height Fe islands 
cannot be  taken 
into account by this kind of PED analysis, but this hypothesis can be 
excluded on the basis of the analysis of the HAS data.

\begin{figure}[tbp]
\begin{center}
\includegraphics[width=.45\textwidth]{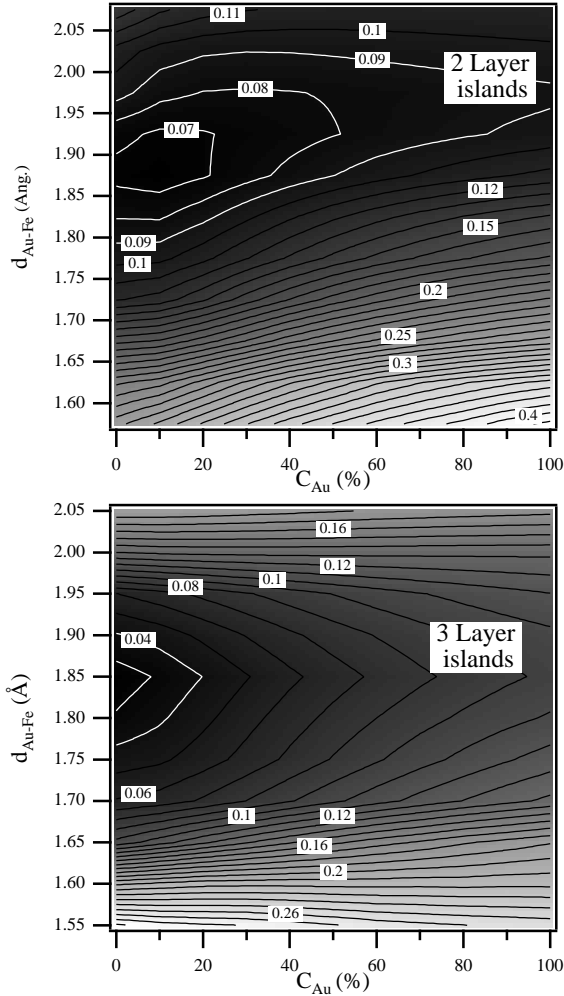}
\end{center}
\caption{R--factor analysis for different extent of Au segregation. 
In the upper (lower) panel the experimental data of the film 
just after deposition (after annealing) is compared 
with a linear combination of Au--substituted model and an 
all-Fe model (see text). 
The concentration $C_{Au}$ is the weight in \% 
of the Au-substituted model.}
\label{fig4}
\end{figure}

Next we checked for the chemical composition of the Fe islands.
The presence of Au 
atoms on top of the Fe film can be 
revealed since they would yield an overall increase of the Fe PED pattern anisotropy, 
due to the Au atom 
cross section which is larger than the Fe atom one.
In the simulated best--fit models we substituted 
the Fe atoms of the topmost layer with a full layer of 
Au atoms and calculated the 
corresponding $\chi$-functions. The resulting $\chi$-functions 
of each Au--substituted film were 
then combined with the $\chi$-functions of the 
corresponding all--Fe--model, using
eq.~\ref{eq2}, where $C=C_{Au}$ now is the concentration of the
Au--substituted film.
The R--factor as a function of $C_{Au}$ and of the height $d_{Au-Fe}$ 
of the Au layer on top of the Fe one is reported in Fig.~\ref{fig4}. 
As can be seen, no Au segregation can be inferred by the PED analysis 
not even after the annealing, therefore the main effect 
of the annealing process is to favor the atoms on the surface to form 
thicker 
islands with the same fcc structure. 

Our R-factor analysis cannnot be extended to check for Cu segregation, 
since the Cu scattering cross-section is close to the Fe one.
On the other hand, Cu segregation is known to be negligible for the 
Fe/Cu(100) system at LT,\cite{chambliss,kief} while 
a full Au layer was observed to segregate on top of Fe films grown on 
Au(100) for many Fe layers.\cite{kellar} Therefore 
the segregation of Cu atoms seems unlikely when the segregation of Au is 
inhibited.

Beyond the chemical composition of Fe nanoislands, we remark the 
different film morphology that can be achieved as a function of 
the substrate temperature during and after the growth. The LT deposition 
of 1~ML of Fe yields a distribution of homogeneously sized bi--layer 
nanoislands, whereas multilayer growth has been reported at RT for a
similar coverage, with the first layer close to percolation limit and 
several third layer islands.\cite{lin2} 
The magnetic measurements reported in the literature are also 
different between films grown at RT\cite{lin,terreni,lin2} and films 
grown at LT, with or without annealing.\cite{feldmann,schirmer}
A magnetization oriented perpendicularly to the surface is 
reported for Fe films, obtained by RT annealing 1~ML of Fe deposited at 
LT.\cite{lin,feldmann} A larger Fe amount is always required for the 
magnetization onset, when the film is grown at RT.\cite{lin,feldmann}
A very similar magnetic 
behavior is reported for thin Fe films grown on Cu(001).\cite{li} 
In the latter case, the onset of the
magnetization (normal to the surface) occurs 
at a critical Fe coverage, which decreases
 as the substrate deposition temperature is 
decreased. A minimum critical coverage of 1 layer equivalent was 
eventually reached upon deposition at 190~K,\cite{li} where only bi--layer 
islands are observed,\cite{giergiel} 
like in the Fe/Cu$_3$Au system.   
We can conclude that the 
onset of  the perpendicular magnetization cannot be simply related to 
the formation of an extended and connected single Fe layer on the 
substrate (as observed for Fe on W(110)\cite{elmers}), 
but rather to the formation of an Fe bulky volume. 
Due to the scattering of data reported in the literature, it is not 
possible to rigorously state the onset of magnetization before the 
coalescence of LT bi--layer islands sets in. 
However, we remark that, even if the magnetization onset is 
delayed up to the bi--layer percolation limit,  the substrate Fermi electrons 
must play a fundamental role  in driving the coherent orientation of 
the Fe island spin (similarly to the magnetic exchange coupling of layered 
systems\cite{ortega}), because of the intrinsic inhomogeinity of the coalescence 
process.

The LT morphology cannot be described by a kinetic model based on the 
interplay between 
intra- and inter-layer diffusion coefficients. 
The Schwoebel barrier for the interlayer 
diffusion is hardly overcome at RT.\cite{stroscio} 
At LT the Fe atoms landing on top of the first layer are 
efficiently trapped in the second layer, but those landing on top of 
the second layer never form a third one, rather they are incorporated 
in the second layer leading to a flat bi--layer island morphology.
A self-induced magnetostriction effect\cite{lin2} can be excluded on 
the basis of energetics arguments. If the kinetic barrier for the atom 
incorporation is of the order of the Schwoebel barrier (i.e. of the 
order of eV \cite{stroscio}), the nanoisland strength of magnetostriction 
should be of many meV/atom, which is almost two order of magnitude 
larger than typical magnetostriction forces.\cite{pbruno}

Very recently, the formation of bi--layer Co islands on Cu(111) was 
explained by a theoretical model predicting the collapse of single 
layer islands above a critical size due to the interplay between the 
excess of strain and exchange processes.\cite{gomez} In fact, the excess of 
Fe surface energy is  lower when considering the (100) surface of both Cu 
and Cu$_{3}$Au.\cite{vitos}  
Moreover, segregation, which would relieve the surface strain to 
stabilize the island growth, is shown to be absent, 
when the bi--layer nanoislands are formed at LT. 

The additional energy excess required to stabilize the Fe nanoislands 
on Cu$_{3}$Au might arise from the quantum confinement of the Fe 
$d$-electrons, like in the model of LT electronic growth proposed by Zhang and 
coworkers.\cite{zhang} In this mean 
field model, islands of ``magical'' thickness are formed due to 
the balance between the stress at the interface, the charge 
transfer and the quantum size effects (QSE). While typical of metals grown
at LT on semiconductors, structural QSE have been also reported for Pb on 
Cu(111).\cite{braun}
Finally, the manifestation of electronic 
phenomena in the low temperature phases of Fe on Cu$_{3}$Au 
seems consistent with the thickness induced oscillations 
($\sim$~3.7~\AA~ period) of the
magnetization observed on the Fe/Cu(100) system at LT.\cite{li} 
These oscillations were in fact related to 
the quantum well states in the Fe film,\cite{ortega} 
i.e. to the manifestation of QSE.

L. F. is grateful to G. Rossi and C. Carbone for useful discussions.
Funding from INFM and from the Italian Ministero dell'Universit\`{a} e
Ricerca Scientifica (Cofin99 Prot. 990211848 and 9902112831) 
are gratefully acknowledged.

%\newpage
%\twocolumn

%\newpage


\begin{thebibliography}{99}

\bibitem{moroni} E.G. Moroni, G. Kresse, J. Hafner, and J. 
Furthm\"uller, Phys. Rev. B {\bf 56}, 15629 (1997).

\bibitem{moruzzi}  V.L. Moruzzi, P.M. Marcus, and J. K\"{u}bler, Phys. Rev.
B {\bf 39}, 6957 (1989); P.M. Marcus, V.L. Moruzzi, and S.L. Qiu, Phys. Rev. B 
{\bf 60 }, 369 (1999).

\bibitem{rochow}  R. Rochow {\it et al.}, Phys. Rev. B {\bf 41}, 3426 (1990).

\bibitem{lin}  M.-T. Lin {\it et al.}, Phys. Rev. B. {\bf 55}, 5886 (1997).

\bibitem{feldmann}  B. Feldmann, B. Schirmer, A. Sokoll, and M. Wuttig, Phys.
Rev. B. {\bf 57}, 1014 (1998).

\bibitem{terreni}  F. Bruno, S. Terreni, L. Floreano, D. Cvetko, P. 
Luches, L. Mattera, A. Morgante, R. Moroni, A. Verdini, and M. Canepa, 
to be published (http://arXiv.org/abs/cond-mat/0103458); F. Bruno 
{\it et 
al.}, Appl. Surf. Sci. {\bf 162-163}, 340 (2000).

\bibitem{luches}  P. Luches, A. Di Bona, S. Valeri, and M. Canepa, Surf. Sci.
{\bf 471}, 32 (2000).

\bibitem{lin2}  M.-T. Lin {\it et al.}, Surf. Sci. {\bf 410}, 290 (1998).

\bibitem{schirmer}  B. Schirmer, B. Feldmann, and M. Wuttig, Phys. Rev. B. 
{\bf 58}, 4984 (1998).

\bibitem{canepa}  M. Canepa, P. Cantini, C. Mannori, S. Terreni, and L.
Mattera, Phys. Rev. B {\bf 62}, 13121 (2000).

\bibitem{chambliss}  K.E. Johnson, D.D. Chambliss, R.J. Wilson, and S. Chiang, 
Surf. Sci. {\bf 313}, L811 (1994).

\bibitem{giergiel} J. Giergiel, J. Shen, J. Woltersdorf, A. Kirilyuk, 
and J. Kirschner, Phys. Rev. B {\bf 52}, 8528 (1995).

\bibitem{figuera} J. de la Figuera, J.E. Prieto, C. Ocal, and R. 
Miranda, Phys. Rev. B {\bf 
47}, 13043 (1993).

\bibitem{gomez} L. G\'{o}mez {\it et al.}, Phys. Rev. Lett. {\bf 84}, 
4397 (2000).

\bibitem{floreano}  L. Floreano {\it et al.}, Rev. of Sci. Inst.{\bf 70}, 3855 
(1999); R. Gotter {\it et al.}, Nucl. Instrum. Methods A {\bf 
467-468}, 1468 (2001);
an updated presentation of the beamline can be found at
http: //tasc.area.trieste.it/ tasc/ lds/ aloisa/ aloisa.html.

\bibitem{mannori}  C. Mannori {\it et al.}, 
Surf. Sci, {\bf 433}, 307  (1999); C. Mannori {\it et al.}, 
Europhysics Lett. {\bf 45}, 686 (1999).

\bibitem{MSCD}  MSCD is freely distributed by Y. Chen and M.A. Van Hove at:
``http://electron.lbl.gov/mscdpack/mscdpack.html''; also see Y. Chen 
{\it et al.}, Phys. Rev. B {\bf 58}, 13121 (1998).

\bibitem{kief} M.T. Kief and W.F. Egelhoff, Phys. Rev. B {\bf 47}, 
10785 (1993).

\bibitem{kellar} S.A. Kellar {\it et al.}, Phys. Rev. B {\bf 57} 1890 (1998); 
R. Opitz, S. L\"{o}bus, A. Thissen, and R. Courths, Surf.
Sci. {\bf 370}, 293 (1997).

\bibitem{li} Dongqi Li, M. Freitag, J. Pearson, Z.Q. Qiu, and S.D. 
Bader, Phys. Rev. Lett. {\bf 72}, 3112 (1994).

\bibitem{elmers} H.J. Elmers {\it et al.}, Phys. Rev. Lett. {\bf 73}, 
898 (1994).

\bibitem{ortega} J.E. Ortega, F.J. Himpsel, G.J. Mankey, and R.F. Willis, Phys. 
Rev. B {\bf 47}, 1540 (1993); M.D. Stiles, Phys. Rev. B {\bf 48}, 
7238 (1993).

\bibitem{stroscio}  J.A. Stroscio, D.T. Pierce, and R.A. Dragoset, Phys. Rev.
Lett. {\bf 70}, 3615 (1993).

\bibitem{pbruno} P. Bruno, {\it Physical origins and theoretical 
models of magnetic anisotropy}, in the proceedings of the 24th 
IFF-Ferienkurs/1993 on {\it Magnetismus von Festk\"orpern und 
Grenzfl\"achen}, p. 24.1, Forschungszentrum J\"ulich GmbH (1993).

\bibitem{vitos} L. Vitos, A.V. Ruban, H.L. Skriver, and J. Kollar, Surf. 
Sci. {\bf 411}, 186 (1998); Ch.E. Leka, N.I. Papanicolaou, and G.A. 
Evangelakis, Surf. Sci. {\bf 479}, 287 (2001).

\bibitem{zhang}  Z. Zhang, Q. Niu, and C.-K. Shih, Phys. Rev. Lett. 
{\bf 80}, 5381 (1998); K.-J. Jin, G.D. Mahan, H. Metiu, and Z. Zhang, 
Phys. Rev. Lett. {\bf 80}, 1026 (1998).

\bibitem{braun} J. Braun and J.P. Toennies, Surf. Sci. Lett. {\bf 
384}, L858 (1997).


\end{thebibliography}
\end{document}